# Throughput and Energy-Efficient Network Slicing

Bho Matthiesen, Osman Aydin and Eduard A. Jorswieck
Institute of Communication Technology, Technische Universität Dresden, Germany
Email: {bho.matthiesen, eduard.jorswieck}@tu-dresden.de

*Abstract*—Network slicing allows 5G network operators to provide service to multiple tenants with diverging service requirements. This paper considers network slicing aware optimal resource allocation in terms of throughput and energy efficiency. We define a heterogeneous Quality of Service (QoS) framework for a sliced radio access network with per-slice zero-forcing beamforming and jointly optimize power and bandwidth allocation across slices and users. The Pareto boundary of this multi-objective optimization problem is obtained by two different algorithms based on the utility profile and scalarization approaches combined with generalized fractional programming. Numerical results show the merits of jointly allocating bandwidth and transmission power and how throughput and global energy efficiency are influenced by slice specific QoS requirements.

*Index Terms*—Resource allocation, multi-objective optimization, Pareto boundary, convex optimization, 5G networks

## I. Introduction

Network slicing is one of the most disruptive new technologies introduced for 5G wireless networks. It allows network operators to provide multiple virtual networks with heterogeneous service requirements on top of a common shared physical infrastructure. The greater elasticity provided by the slicing concept combined with network function virtualisation has been identified as a key technology for efficient 5G network resource usage by the Next Generation Mobile Networks (NGMN) consortium [1], [2]. Another key technology for 5G networks enabling sustainable growth is energy-efficient resource allocation. 5G networks are expected to increase the data rate by a factor of 1000 while, at the same time, reducing the power consumption by a factor of 2. Thus, a 2000x increase of the bit-per-Joule energy efficiency (EE) is required [1], [3]. A recent survey covering the main approaches to make 5G energy-efficient is [4]. In line with these observations on current research trends for 5G and the requirements from NGMN [1], we consider throughput (TP) and energy-efficient resource allocation under heterogeneous Quality of Service (QoS) constraints.

A key aspect of network slicing is the physical isolation across slices which is achieved by time or frequency orthogonality. While the time domain is typically handled by the scheduler, both the available bandwidth and transmit power at the base station can be distributed by the resource allocation algorithm. In general, four different combinations of bandwidth

This work is supported in part by the German Research Foundation (DFG) in the Collaborative Research Center 912 "Highly Adaptive Energy-Efficient Computing".

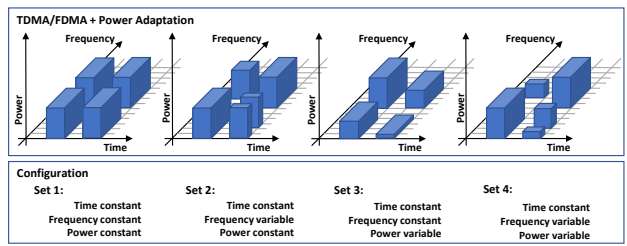

Fig. 1. The four combination possibilities for dynamic power and sub-carrier assignments

and transmission power allocations are possible. These are outlined in Fig. 1 and described below:

1) Set I describes static distribution of both, bandwidth and transmission power resources. In this scenario the focus is to achieve given QoS requirements without any trade-offs. Data throughput control per user is possible by appropriate assignment of scheduling time.
2) Set II describes the general implementation in wireless networks. TDMA and FDMA techniques are combined in order to achieve QoS requirements. UEs are scheduled with a certain number of frequency resources with a fixed average transmission power.
3) Set III is described as Capacity Adaptation (CAP) approach in [5]: CAP is a method, which does not change the maximum used bandwidth and the number of reference signals. An adaptation to lower load is performed by scheduling only a part of the subcarriers, i.e. limiting the number of scheduled physical resource blocks (PRBs). This approach allows lowering the power amplifier (PA) supply voltage, but it is transparent to the mobile terminals and maintains frequency diversity. CAP is using a reconfiguration of cell parameters, is very slow and may impact the performance of cell-edge users.
4) Set IV is described as Bandwidth Adaptation (BW) approach in [5]: BW is based on the adjustment of the bandwidth to the required traffic load. Depending on traffic load the bandwidth can be stepwise downscaled that lower numbers of PRBs are allocated. Then the PA can adapt to lower supply voltage and less reference signals have to be sent. [5]

In this paper, network slicing aware joint power and bandwidth allocation under heterogeneous QoS constraints is considered. Sets I–III as defined above are treated as special

cases. The contributions of this paper can be summarized as follows:

- We model a network slicing downlink system and state the resource allocation problem with heterogeneous QoS constraints as a multi-objective optimization problem (MOP) of the two most widely used physical layer performance metrics: energy efficiency and throughput. We show that the feasible set is jointly convex in the bandwidth and power allocation and discuss the treatment of physical resource blocks.
- Two solution approaches for this MOP are discussed: The utility profile and the scalarization approach. We show that the utility profile approach is a generalized fractional program that can be solved as a sequence of convex programs by the generalized Dinkelbach's algorithm with polynomial complexity. Instead, the scalarization approach results in a sum-of-ratios problem and belongs to the class of global optimization problems. We propose a one dimensional line search algorithm that finds the global optimal solution with affordable complexity.
- We evaluate the performance of both algorithms numerically and illustrate the gain of joint power and bandwidth allocation over fixed resource allocation.

## II. Sliced Network System Model

We consider an OFDM-based multi-user multiple-input single-output (MISO) downlink transmission system. We focus on the coverage area of a single $M$ antenna base station (BS) with total bandwidth $B$ and maximum transmit power $P$. One or more network slices $\mathcal{S}_i$, $i = 1, \ldots, I$, with bandwidth $b_i$ are instantiated dynamically at the BS. It serves $K$ active single antenna terminals each belonging to exactly one slice.[1]

Let $s_k$ be the index of user $k$'s slice, i.e., it is chosen such that $k \in \mathcal{S}_{s_k}$. Then, the received signal at user $k$ is

$$y_k = \sqrt{p_k} \boldsymbol{h}_k^H \boldsymbol{w}_k x_k + \sum_{\substack{i \in \mathcal{S}_{s_k} \\ i \neq k}} \sqrt{p_i} \boldsymbol{h}_k^H \boldsymbol{w}_i x_i + z_k,$$

where $p_k$ is the transmit power allocated for user $k$, $\boldsymbol{h}_k$ is the $M \times 1$ channel vector between the BS and user $k$, $\boldsymbol{w}_k$ is the unit-norm beamforming vector for user $k$, $x_k$ is the unit-variance information symbol intended for user $k$ and $z_k$ is the noise observed at receiver $k$, modelled as circularly symmetric complex Gaussian noise with power $N_0 b_{s_k}$.

The BS employs zero-forcing (ZF) beamforming for each slice. This limits the maximum number of users served per slice to the number of transmit antennas, i.e., $\max_i |\mathcal{S}_i| \leq M$. The downlink channel in each slice $\mathcal{S}_i$ is formally equivalent to an $|\mathcal{S}_i| \times M$ multiple-input multiple-output (MIMO) system with channel matrix $\boldsymbol{H}_i = [\boldsymbol{h}_k]_{k \in \mathcal{S}_i}^H \in \mathbb{C}^{|\mathcal{S}_i| \times M}$. Thus, the beamformers for slice $\mathcal{S}_i$ are $\boldsymbol{w}_k = \frac{\tilde{\boldsymbol{w}}_k}{\|\tilde{\boldsymbol{w}}_k\|}$, $k \in \mathcal{S}_i$, with $[\tilde{\boldsymbol{w}}_k]_{k \in \mathcal{S}_i} = \boldsymbol{H}_i^H (\boldsymbol{H}_i \boldsymbol{H}_i^H)^{-1}$, and the achievable rate at user $k$ is

$$R_k(b_{s_k}, p_k) = b_{s_k} \log(1 + \gamma_k), \quad (1)$$

where

$$\gamma_k = \frac{\left|\boldsymbol{h}_k^H \boldsymbol{w}_k\right|^2 p_k}{N_0 b_{s_k}} \quad (2)$$

is the receive signal to interference plus noise ratio (SINR). Then, the TP of slice $\mathcal{S}_i$ and the system TP, respectively, are defined as

$$\text{TP}_i(\boldsymbol{b}, \boldsymbol{p}) = \sum_{k \in \mathcal{S}_i} b_i \log(1 + \gamma_k), \quad (3)$$

$$\text{TP}(\boldsymbol{b}, \boldsymbol{p}) = \sum_{k=1}^{K} b_{s_k} \log(1 + \gamma_k), \quad (4)$$

where $\boldsymbol{b} = (b_1, b_2, \ldots, b_I)$ and $\boldsymbol{p} = (p_1, p_2, \ldots, p_K)$. The system TP measures the TP from the network slicing provider's perspective, while the slice TP is the relevant TP metric for the slice tenant.

---

[1]In general, a network-slice-aware terminal might be capable of subscribing to multiple network slice instances simultaneously. From the resource allocation perspective this can be handled without loss of generality by adding several virtual terminals.

Similarly, the bit-per-Joule EE of user $k$, the EE of slice $\mathcal{S}_i$ (SEE) and the system global energy efficiency (GEE) are defined as [6]

$$\text{EE}_k(b_{s_k}, p_k) = \frac{b_{s_k} \log(1 + \gamma_k)}{P_{0,k} + \phi_k p_k}, \quad (5)$$

$$\text{SEE}_i(\boldsymbol{b}, \boldsymbol{p}) = \frac{\sum_{k \in \mathcal{S}_i} b_i \log(1 + \gamma_k)}{P_{0,\mathcal{S}_i} + \sum_{k \in \mathcal{S}_i} \phi_k p_k}, \quad (6)$$

$$\text{GEE}(\boldsymbol{b}, \boldsymbol{p}) = \frac{\sum_{k=1}^{K} b_{s_k} \log(1 + \gamma_k)}{P_0 + \sum_{k=1}^{K} \phi_k p_k}, \quad (7)$$

where the positive constants $\phi_k$ and $P_{0,k}$ model the power amplifiers inefficiency and static circuit power consumption of user $k$, respectively, and $P_{0,\mathcal{S}_i} = \sum_{k \in \mathcal{S}_i} P_{0,k}$ and $P_0 = \sum_{k=1}^{K} P_{0,k}$ are slice $\mathcal{S}_l$'s and the system's total circuit power consumption, respectively.

User selection is beyond the scope of this paper and assumed to be done by a higher layer authority prior to resource allocation. It can be done either by the network slicing provider or the tenant. In any case, it must be slicing aware, meaning that users are not to be scheduled across slices since they are subscribed to a specific service hosted within the network slice instance. Moreover, the scheduler may need to take some QoS constraints of the slice into account to ensure, e.g., sufficient average throughput or latency constraints. Of course, the scheduler must also be channel aware to operate efficiently. Considering the ZF beamforming employed here, a strong candidate is the semi-orthogonal user selection algorithm [7], applied on a per slice basis. However, it might be necessary to extend it to incorporate QoS constraints.

## III. Optimal Resource Allocation

The network slicing provider is usually interested in maximizing both the system TP and GEE to utilize the available bandwidth as best as possible and keep operating costs at a minimum. In general, the TP and GEE are conflicting performance measures, since the maximization of one can lead to a decrease of the other. Thus, we formulate the program to be solved as the MOP [8], [9]

$$\max_{(\boldsymbol{b},\boldsymbol{p})\in\mathcal{D}} [\text{GEE}(\boldsymbol{b},\boldsymbol{p}), \text{TP}(\boldsymbol{b},\boldsymbol{p})] \quad \text{(P1)}$$

where $\mathcal{D}$ is the feasible set defined by the following constraints:

- physical constraints and the BS's resource limits:

$$\boldsymbol{b} \geq 0, \quad \boldsymbol{p} \geq 0 \quad (8a)$$

$$\sum_{i=1}^{I} b_i \leq B, \quad \sum_{k=1}^{K} p_k \leq P \quad (8b)$$

- QoS constraints of the network slicing provider:

$$\text{GEE}(\boldsymbol{b},\boldsymbol{p}) \geq \text{GEE}^\star, \quad \text{TP}(\boldsymbol{b},\boldsymbol{p}) \geq \text{TP}^\star \quad (8c)$$

- QoS constraints of slice $\mathcal{S}_i$:

$$\text{TP}_i(\boldsymbol{b},\boldsymbol{p}) \geq \text{TP}_i^\star, \quad \text{SEE}_i(\boldsymbol{b},\boldsymbol{p}) \geq \text{SEE}_i^\star \quad (8d)$$

- resource limits for slice $\mathcal{S}_i$:

$$\sum_{k\in\mathcal{S}_i} p_k \leq P_i, \quad b_i \leq B_i \quad (8e)$$

- QoS constraints of user $k$:

$$R_k(b_{s_k}, p_k) \geq R_k^\star, \quad \text{EE}_k(b_{s_k}, p_k) \geq \text{EE}_k^\star \quad (8f)$$

where the starred values are QoS related constants negotiated in advance between the network slicing provider and the tenants.

Commonly cited examples of network slices are mobile broadband with very moderate QoS constraints, Internet of Things (IoT) with strong EE requirements, and very low latency applications like vehicular communications [1]. A simple resource allocation approach to achieving low latency communication is keeping the queueing delay down, which is equivalent to minimum rate constraints.

*Lemma 1:* The feasible set $\mathcal{D}$ defined by (8a)–(8f) is closed and convex.

*Proof:* First, observe that the function $b_{s_k}\log(1+\gamma_k)$ is jointly concave in $b_{s_k}$ and $p_k$. Since the sum of those functions is also concave, the rate and TP constraints are convex constraints.

Consider the GEE constraint in (8c). With (7) this is equivalent to

$$\text{GEE}^\star\left(P_0 + \sum_{k=1}^{K}\phi_k p_k\right) - \sum_{k=1}^{K} b_{s_k}\log(1+\gamma_k) \leq 0. \quad (9)$$

Since the left-hand side (LHS) is a convex function, this is a convex constraint. Similarly, (8d) and (8f) are equivalent to

$$\text{SEE}_i^\star\left(P_{0,\mathcal{S}_i} + \sum_{k\in\mathcal{S}_i}\phi_k p_k\right) - \sum_{k\in\mathcal{S}_i} b_i\log(1+\gamma_k) \leq 0 \quad (10)$$

$$\text{EE}_k^\star(P_{0,k} + \phi_k p_k) - b_{s_k}\log(1+\gamma_k) \leq 0 \quad (11)$$

and, thus, convex constraints.

Finally, since (8a), (8b) and (8e) are affine and none of the inequalities is strict, $\mathcal{C}$ is a closed convex set. ∎

From the perspective of optimization, it is well-known that the MOP (P1) usually admits infinite number of noninferior solutions, which form the outermost boundary of the achievable performance region, the so-called Pareto boundary [8]. More specifically, the achievable region is defined as the set of all feasible pairs $(\text{GEE}(\boldsymbol{b},\boldsymbol{p}), \text{TP}(\boldsymbol{b},\boldsymbol{p}))$, i.e.,

$$\mathcal{P} = \{(\text{GEE}(\boldsymbol{b},\boldsymbol{p}), \text{TP}(\boldsymbol{b},\boldsymbol{p})) : (\boldsymbol{b},\boldsymbol{p}) \in \mathcal{D}\}. \quad (12)$$

The outer frontier of $\mathcal{P}$ then is the Pareto boundary $\partial^+\mathcal{P}$ of (P1) defined as

$$\partial^+\mathcal{P} = \{\boldsymbol{x} \in \mathcal{P} | \nexists \boldsymbol{x}' \in \mathcal{P} : (\forall i : x_i' \geq x_i) \wedge (\exists i : x_i' > x_i)\} \quad (13)$$

and represents the set of all Pareto-optimal points. All Pareto-optimal points have the property that it is impossible to increase one of the objectives without decreasing the other.

### A. Special Cases: Sets I–III

Recall from Fig. 1 the four different possibilities for dynamic power and sub-carrier allocation. The above discussion treats the general case of variable power and frequency denoted as Set IV. The other sets are special cases easily obtained from (P1). Consider the case of constant power allocation in Set II. Here, each user $k$ gets a fixed fraction $\rho_k$ of the total allocated power $\mathfrak{P}$, i.e., $\boldsymbol{p} = \mathfrak{P}\boldsymbol{\rho}$, where $\rho_k > 0$ and $\sum_{k=1}^{K}\rho_k = 1$. Then, the transmit power constraint in (8b) becomes $\mathfrak{P} \leq P$ and the optimization in (P1) is over $(\boldsymbol{b}, \mathfrak{P})$. Similarly, for the fixed bandwidth allocation in Set III each slice $i$ gets a fixed ratio $\beta_i$ of the total allocated bandwidth $\mathfrak{B}$, i.e., $\boldsymbol{b} = \mathfrak{B}\boldsymbol{\beta}$, with $\beta_i > 0$ and $\sum_{i=1}^{I}\beta_i = 1$. The bandwidth constraint in (8b) then is $\mathfrak{B} \leq B$ and the optimization in (P1) is over $(\mathfrak{B}, \boldsymbol{p})$. Finally, Set I is a combination of Sets II and III where the optimization in (P1) is over $(\mathfrak{B}, \mathfrak{P})$. The solution approaches to (P1) presented in the next section are also valid for these special cases with the minor modifications mentioned above. Thus, we do not treat them explicitly.

### B. Physical Resource Blocks

The assumption above is that bandwidth can be allocated in arbitrarily sized chunks. In a real world setting, the bandwidth is rather an integer multiple of the PRB size. Given that the PRB size is small enough compared to the total bandwidth $B$, that might be a valid assumption. However, the PRB size $B_{\text{PRB}}$ can be taken explicitly into account by a simple modification of the feasible set $\mathcal{D}$. For each $b_k$, add a nonnegative integer variable $\xi_k$ and the condition $b_k = \xi_k B_{\text{PRB}}$.

Of course, the feasible set is then no longer convex. Instead, the resulting optimization problems considered in the next section then belong to the class of mixed integer programming problems and have, in general, exponential complexity. For fixed integer variables, the feasible set is still convex and exploiting this fact allows for an efficient implementation. Moreover, mixed integer disciplined convex programming (MIDCP) is applicable, but, especially for a large number of

network slices, a solution might not be obtainable in reasonable time.[2]

## IV. CHARACTERIZATION OF THE PARETO BOUNDARY

Multi-objective programming theory provides several approaches to convert the vector objective of a MOP into a scalar one whose maximization results in a Pareto-optimal point. The two most widely used are the utility profile approach and scalarization [11]. The utility profile approach finds the intersection of a ray and the Pareto boundary. By varying the direction of the ray, the complete Pareto boundary can be characterized. Instead, in the scalarization approach, the weighted sum of the objectives is maximized. In this case, by varying the weights, the convex hull of the Pareto boundary is characterized. This, however, is not necessarily a limitation since time sharing between Pareto-optimal points allows to achieve every point on the convex hull of the Pareto boundary.

### A. Utility Profile Approach

The intersection of the Pareto boundary and a ray in the direction of $(\alpha, 1-\alpha)$, $0 \leq \alpha \leq 1$, is the result of the optimization problem

$$\begin{cases} \max_{t,\bm{b},\bm{p}} & t \\ \text{s.t.} & \text{GEE}(\bm{b},\bm{p}) \geq \alpha t \\ & \text{TP}(\bm{b},\bm{p}) \geq (1-\alpha)t \\ & (\bm{b},\bm{p}) \in \mathcal{D}. \end{cases} \quad \text{(P2)}$$

This is a non-convex optimization problem because the right-hand side (RHS) of the GEE constraint, which is equivalent to

$$\sum_{k=1}^{K} b_{s_k} \log(1+\gamma_k) \geq \alpha t \left( P_0 + \sum_{k=1}^{K} \phi_k p_k \right), \quad (14)$$

is not jointly convex in $(t, \bm{p})$. Instead, we reformulate (P2) as

$$\max_{(\bm{b},\bm{p}) \in \mathcal{D}} \min \left( \frac{1}{\alpha} \frac{\sum_{k=1}^{K} b_{s_k} \log(1+\gamma_k)}{P_0 + \sum_{k=1}^{K} \phi_k p_k}, \right.$$
$$\left. \frac{1}{1-\alpha} \sum_{k=1}^{K} b_{s_k} \log(1+\gamma_k) \right), \quad \text{(P3)}$$

which is a generalized fractional program, and known to be quasiconcave provided each ratio has a concave numerator and convex denominator [6, Sec. 2.3.4 & 3.3].

A fundamental result of generalized fractional programming establishes that solving (P3) is equivalent to finding the unique root of $\max_{(\bm{b},\bm{p}) \in \mathcal{D}} F(\bm{b},\bm{p};\lambda)$, where

$$F(\bm{b},\bm{p};\lambda) = \min \left\{ \frac{1}{1-\alpha} \sum_{k=1}^{K} b_{s_k} \log(1+\gamma_k) - \lambda, \right.$$
$$\left. \frac{1}{\alpha} \sum_{k=1}^{K} b_{s_k} \log(1+\gamma_k) - \lambda \left( P_0 + \sum_{k=1}^{K} \phi_k p_k \right) \right\}. \quad (15)$$

[2]For example, the popular disciplined convex programming toolbox CVX [10] implements MIDCP since version 2.0.

This is accomplished by the generalized Dinkelbach's algorithm (GDA) which is stated in Algorithm 1. Observe that the inner problem (16) is a convex optimization problem and, thus, solvable with polynomial complexity, while the outer algorithm has linear convergence rate [12].

---
**Algorithm 1** Generalized Dinkelbach's algorithm [12]
---
Initialize $\varepsilon > 0$, $j = 0$, $(\bm{b}^0, \bm{p}^0) \in \mathcal{D}$.
**repeat**
$\quad j \leftarrow j+1$
$\quad \lambda^j \leftarrow \min \left\{ \text{GEE}(\bm{b}^{j-1}, \bm{p}^{j-1}), \text{TP}(\bm{b}^{j-1}, \bm{p}^{j-1}) \right\}$
$\quad (\bm{b}^j, \bm{p}^j) \leftarrow \arg\max_{(\bm{b},\bm{p}) \in \mathcal{D}} F(\bm{b},\bm{p};\lambda^j) \quad (16)$
**until** $F(\bm{b}^j, \bm{p}^j; \lambda^j) \leq \varepsilon$

---

### B. Scalarization Approach

In the scalarization approach, the weighted sum of the objectives is maximized, i.e.,

$$\max_{(\bm{b},\bm{p}) \in \mathcal{D}} \alpha \text{GEE}(\bm{b},\bm{p}) + (1-\alpha)\text{TP}(\bm{b},\bm{p}). \quad \text{(P4)}$$

This optimization problem is a slightly degenerate instance of the sum-of-ratios problems and is significantly harder to solve than (P2). The general problem of maximizing $f(x) + \frac{g(x)}{h(x)}$ with $f$, $g$, and $-h$ concave functions, and $g(x) \geq 0$ and $h(x) > 0$ is known to be essentially $\mathcal{NP}$-complete [13]. Indeed, if we recast Problem (P4) as

$$\begin{cases} \max_{t,\bm{b},\bm{p}} & \alpha \dfrac{t}{P_0 + \sum_{k=1}^{K} \phi_k p_k} + (1-\alpha)t \\ \text{s.t.} & \sum_{k=1}^{K} b_{s_k} \log(1+\gamma_k) \geq t \\ & (\bm{b},\bm{p}) \in \mathcal{D}, \end{cases} \quad \text{(P5)}$$

we can prove the following lemma.

*Lemma 2:* The objective of (P5) is pseudoconvex.

*Proof:* First, note that the objective is equivalent to $t \left( \alpha \frac{1}{P_0 + \sum_{k=1}^{K} \phi_k p_k} + 1 - \alpha \right) := t\Phi(\bm{p})$. Observe that $\Phi(\bm{p})$ is a convex function and that $1/\Phi(\bm{p})$ is concave. Since both are differentiable, $t \geq 0$, and $\Phi(\bm{p}) > 0$ their product $t\Phi(\bm{p})$ is a pseudoconvex function [14, Table 5.2]. ∎

Lemma 2 opens the possibility to use the vast toolbox of (pseudo-)concave minimization theory; see e.g. [15]. However, for fixed $t$, (P5) is equivalent to minimizing $\sum_{k=1}^{K} \phi_k p_k$ over (P5)'s feasible set, i.e.

$$\min_{(\bm{b},\bm{p}) \in \mathcal{D}} \sum_{k=1}^{K} \phi_k p_k \quad \text{s.t.} \quad \sum_{k=1}^{K} b_{s_k} \log(1+\gamma_k) \geq t. \quad \text{(P6)}$$

Then, the solution of (P5) can be obtained by performing a line search over $t$, solving this convex subproblem (P6) for each $t$ [16]. In order for the line search to be efficient, we need suitable bounds on the range of $t$. The upper bound $\bar{t}$ is the solution to the convex optimization problem $\max_{(\bm{b},\bm{p}) \in \mathcal{D}} \text{TP}(\bm{b},\bm{p})$ and that a lower bound $\underline{t}$ is easily obtained from the QoS constraints as $\max\{\sum_{k=1}^{K} R_k^\star, \sum_{i=1}^{I} \text{TP}_i, \text{TP}\}$. The resulting algorithm is formally stated in Algorithm 2.

**Algorithm 2** Line search algorithm for Problem (P4).
---
Initialize $\underline{t}, \bar{t}, \gamma^\star = -\infty$.
**for** $t \in [\underline{t}, \bar{t}]$ **do**
    Solve (P6). Store solution and optimal value in $(\tilde{\boldsymbol{b}}, \tilde{\boldsymbol{p}})$ and $\tilde{\gamma}$.
    **if** $\tilde{\gamma} \geq \gamma^\star$ **then**
        $\gamma^\star \leftarrow \tilde{\gamma}$
        $(\boldsymbol{b}^\star, \boldsymbol{p}^\star) \leftarrow (\tilde{\boldsymbol{b}}, \tilde{\boldsymbol{p}})$
    **end if**
**end for**

## V. NUMERICAL EVALUATION

For the numerical evaluation, we consider 10 users grouped in 4 slices, where slice $i$ serves $i$ users. The BS has 4 antennas and a maximum transmit power of 10 W. Channels are independent and identically distributed (i.i.d.) zero-mean, unit variance circularly symmetric complex Gaussian, $N_0 = 10^{-2} \frac{\text{W}}{\text{Hz}}$, $P_{0,k} = 0.1 \text{W}$, and $\phi_k = 4$. Initially, we do not consider any QoS constraints.

Figure 2 shows the Pareto boundary of the four combination possibilities for dynamic power and sub-carrier assignment (cf. Fig. 1) for a bandwidth of 100 Hz[3] obtained by both, the utility profile and the scalarization approach. First, observe that jointly optimizing bandwidth and power results in a significantly larger performance region than the other sets. Also, Set III, i.e. fixed bandwidth allocation, achieves a larger region than the remaining two sets that fix the transmit power and both variables, respectively. It can be seen that the performance gain in the GEE domain is higher than in the TP domain which indicates that fixing the power is much more constraining than fixing the bandwidth. Moreover, optimizing both variables jointly results in a much higher performance gain than can be obtained by just optimizing one variable.

As already pointed out in Section IV the scalarization approach finds the convex hull of the Pareto boundary while the utility profile approach directly characterizes $\partial^+ \mathcal{P}$. It is obvious from Fig. 2 that $\partial^+ \mathcal{P}$ is non-convex in general. Depending on the application directly obtaining the convex hull might be desirable or not. However, the scalarization approach requires the solution of a global optimization problem and has, thus, exponential complexity, while the utility profile approach is solvable in polynomial time. Moreover, the convex hull can be obtained at virtually no cost from $\partial^+ \mathcal{P}$. Thus, the utility profile approach and Algorithm 1 should be preferred over the scalarization approach in most cases.

Figure 3 displays a second set of Pareto boundaries obtained by the utility profile approach for a bandwidth of 1000 Hz. The same performance observations as in Fig. 2 can be made. In this case, the performance gap between Set IV and the other sets is even more pronounced. For comparison, the results from Fig. 2 are also shown. Observe that with joint power and bandwidth allocation a higher GEE is achievable for a bandwidth of 100 Hz at almost the same TP than with ten times the bandwidth and limited resource allocation.

---
[3]We have considered small bandwidths due to simulation complexity. The scaling to usual LTE bandwidth of 10, 20 MHz and more can be done in a similar way but requires in particular for the scalarization approach more computational power. Therefore, we restricted the numerical simulations to these smaller bandwidth values.

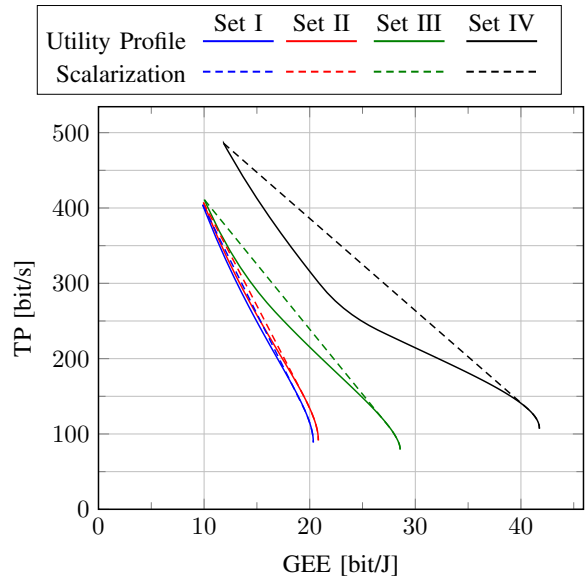

Fig. 2. Pareto boundary for dynamic power and sub-carrier allocation with a bandwidth of 100 Hz: Utility profile versus scalarization approach.

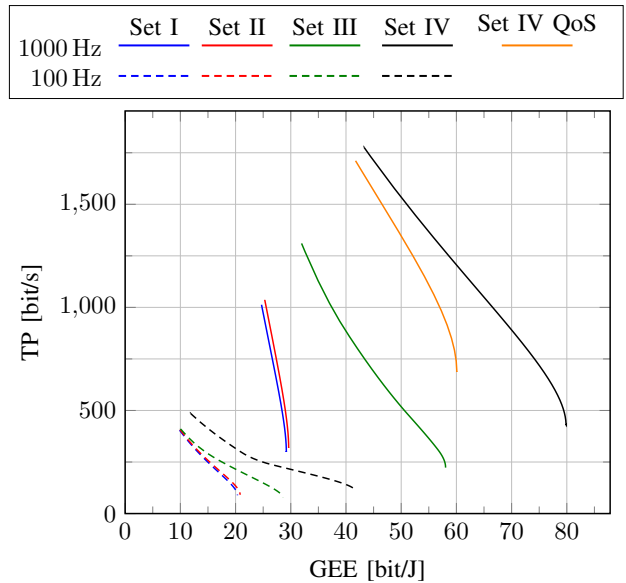

Fig. 3. Pareto boundary obtained by the utility profile approach for bandwidths of 100 Hz and 1000 Hz. The 1000 Hz case is evaluated with and without QoS constraints.

Figure 3 also shows the Pareto boundary for a scenario with QoS constraints. In particular, we assume that the network slicing provider requires a GEE of at least 40 bit/J for cost efficient operation. The users in slice 2 both demand an EE of at least 15 bit/J and the users in slice 4 demand a minimum rate of 5 bit/s. Finally, the tenants of slice 3 and 4 require a SEE of 15 bit/J and a TP of 50 bit/s, respectively. These constraints are only feasible for Set IV and, thus, Fig. 3 only shows the Pareto boundary for this set. As expected, the achievable Pareto region lies within the Pareto region of Set IV without QoS constraints. This demonstrates again the merits of joint power and bandwidth allocation since these constraints are not serviceable in the other sets without using additional resources.

Without QoS constraints, the optimal resource allocation for Set IV usually serves just a single slice. The served slice might change with $\alpha$ and during this change there is a small transition period where two clusters are served. When QoS constraints are present, the optimal resource allocation first fulfills the QoS constraints and then allocates the remaining resources to the strongest slice.

## VI. CONCLUSIONS

We have considered network slicing aware joint bandwidth and power allocation with the goal of maximizing TP and EE. Two algorithms to characterize the Pareto-boundary based on the utility profile and scalarization approaches are proposed. It is shown that the utility profile based algorithm has polynomial complexity and that, compared to more traditional resource allocation schemes, significant gains are achievable thru joint optimization of bandwidth and power allocation.